\documentclass[useAMS,usenatbib]{mn2e}

\title[Li abundance in M54] 
{The cosmological Lithium problem outside the Galaxy: \\
the Sagittarius globular cluster M54
\thanks{Based on data taken at the ESO, within 
the observing program 089.D-0341.}}

\author[Mucciarelli et al.]
  {A. Mucciarelli,$^1$ M. Salaris,$^2$, P. Bonifacio$^3$, 
  L. Monaco$^4$ and S. Villanova$^5$
  \\
  $^1$Dipartimento di Fisica \& Astronomia, Universit\`a 
  degli Studi di Bologna, Viale Berti Pichat, 6/2 - 40127, 
  Bologna, Italy
  \\
  $^2$Astrophysics Research Institute, Liverpool John Moores University, 
    IC2, 146 Brownlow Hill, Liverpool L3 5RF, United Kingdom 
   \\
   $^3$GEPI, Observatoire de Paris, CNRS, Univ. Paris Diderot, 92125, Meudon Cedex, France
   \\
   $^4$European Southern Observatory, Casilla 19001, Santiago, Chile
   \\
   $^5$Universidad de Concepcion, Casilla 160-C, Concepcion, Chile
}

\usepackage{graphicx}

\usepackage{color}

\def\LaTeX{L\kern-.36em\raise.3ex\hbox{a}\kern-.15em
    T\kern-.1667em\lower.7ex\hbox{E}\kern-.125emX}

\begin{document}

\label{firstpage}

\maketitle


\begin{abstract}  
The cosmological Li problem is the observed discrepancy between 
Li abundance ($A(Li)$) measured in Galactic dwarf, old and metal-poor stars (traditionally assumed 
to be equal to the initial value $A(Li)_0$), and that predicted by  
standard Big Bang Nucleosynthesis calculations ($A(Li)_{BBN}$). 
Here we attack the Li problem by considering an alternative diagnostic, namely the surface Li abundance 
of red giant branch stars that in a colour magnitude diagram populate the region 
between the completion of the first dredge-up and the red giant branch bump.
We obtained high-resolution spectra with the FLAMES facility at the Very Large Telescope 
for a sample of red giants in the globular cluster M54, belonging to the Sagittarius dwarf galaxy.  
We obtain $A(Li)$=~0.93$\pm$0.11 dex, translating -- after taking into account the dilution due to the dredge up-- 
to initial abundances ($A(Li)_0$) in the range 2.35--2.29~dex, 
depending on whether or not atomic diffusion is considered. 
This is the first measurement of Li in the Sagittarius galaxy and 
the more distant estimate of $A(Li)_0$ in old stars obtained so far. 
The $A(Li)_0$ estimated in M54 is lower by $\sim$0.35 dex 
than $A(Li)_{BBN}$, hence incompatible at a level of $\sim3\sigma$. 
Our result shows that this discrepancy is 
a universal problem concerning both the Milky Way and extra-galactic systems. 
Either modifications of BBN calculations, or 
a combination of atomic diffusion plus a suitably tuned additional mixing during the main sequence, need to be invoked 
to solve the discrepancy.
\end{abstract}  




\section{Introduction}   
\label{intro}  
Lithium, together with hydrogen and helium, is produced in the first minutes after the Big Bang, 
and its primordial abundance is a function of the cosmological density of baryons. 
An estimate of this primordial Li abundance provides therefore an important test for 
current standard cosmological models.
\citet{spite82} first discovered that dwarf (main sequence, turn-off or sub-giants), Population II stars 
with effective temperatures ($T_{eff}$) between $\sim$5700 and $\sim$6300 K and [Fe/H]$<$--1.4 dex 
share the same Li abundance, the so-called {\sl Spite Plateau}. 
The existence of a narrow Li {\sl Plateau} has been confirmed by three decades of 
observations \citep[see e.g.][]{rebolo88,boni97,asplund06,boni07}; when considering stellar 
evolution calculations that include only convection as element transport, this plateau corresponds 
to the primordial Li abundance in the Galactic halo, that is usually identified as 
the Li abundance produced during the Big Bang Nucleosynthesis ($A(Li)_{BBN}$). 
The measured Li abundance in {\sl Spite Plateau} dwarfs is in the range
A(Li)\footnote{A(Li)=$\log\frac{n(Li)}{n(H)}+12.00$}=~2.1--2.3~dex, 
depending on the adopted $T_{eff}$ scale.

On the other hand, the very accurate determination of the baryonic density obtained from the WMAP 
\citep{spergel07,hinshaw13} and PLANCK \citep{planck} satellites, 
coupled with the BBN standard model, has allowed to calculate $A(Li)_{BBN}$. 
The derived values 
\citep[2.72$\pm$0.06 dex,][and 2.69$\pm$0.04, Coc et al. 2013]{cyburt08}  
are significantly higher, about a factor of 3, than that measured in dwarf stars. 

A first potential solution to this discrepancy between $A(Li)_{BBN}$ from BBN calculations and 
{\sl Spite Plateau} measurements (denoted here as the {\sl cosmological Li problem}) 
envisages the inclusion of atomic diffusion in stellar model calculations. Atomic diffusion is   
a physical process that can be modeled parameter-free from first principles, 
it is efficient in the Sun \citep[see e.g.][]{bahc}, and   
can deplete efficiently the surface abundance of Li in metal poor main sequence stars. However,  
because the degree of depletion increases with effective temperature (and decreasing metallicity), 
it is not possible to 
reproduce the observed plateau-like abundance trend \citep[see e.g.][and references therein]{richard05} 
if atomic diffusion is fully efficient in objects populating the {\sl Spite Plateau}, 
see e.g. Fig.~3 in \citet{m11}.

Recent proposed solutions to the {\sl cosmological Li problem}) are: 
  
{\sl (i)}~the combined effect of atomic diffusion and some competing additional mixing 
--necessary to preserve the existence of an abundance {\sl plateau}-- whose combined effect 
decreases the Li abundance in the atmospheres of dwarf stars 
\citep{richard05,korn06}; 
{\sl (ii)}~inadequacies of the BBN model used to calculate $A(Li)_{BBN}$ \citep[see e.g.][]{iocco09};
{\sl (iii)}~a Li depletion driven by Population III stars during the early Galaxy evolution 
\citep{piau}.

\citet[][MSB12]{msb12} proposed an alternative/complementary route 
to investigate the initial Li abundance in Population II stars ($A(Li)_0$), 
by measuring the surface Li abundance in lower red giant branch (RGB) stars. 
These stars are located between the completion of the 
first dredge-up (FDU, where Li-free material is mixed to the surface by convection) and the 
luminosity level of the RGB bump \citep[where an additional mixing episode occurs, see][]{gratton00}. 
These giants are characterised by a constant Li abundance (at fixed [Fe/H]), 
drawing a {\sl Plateau} that mirrors the {\sl Spite Plateau} but at a lower abundance 
(A(Li)$\sim$0.9-1.0 dex). 
The amount of Li depletion due to dilution after the FDU 
can be predicted easily by stellar models.
Lower RGB stars are therefore a powerful alternative diagnostic of $A(Li)_0$, mainly because 
the derived value is very weakly affected by atomic diffusion 
during the previous main sequence phase. This means that it is possible to put strong constraints 
on $A(Li)_0$, irrespective of whether atomic diffusion is effective or not, and assess whether additional 
processes --within the stars, or during the BBN nucleosynthesis, or during Galaxy formation-- need to be invoked 
to match the BBN calculations of Li abundances.
Moreover, lower RGB stars also enable to investigate
$A(Li)_0$ in stars more distant than those usually observed for  
{\sl Spite Plateau} studies. 

In this paper we exploit this new diagnostic with the aim to study $A(Li)_0$ 
in M54, a massive globular cluster (GC) immersed in the nucleus of the Sagittarius (Sgr) 
dwarf galaxy \citep{monaco05,bellazzini08}. 
The dwarf stars in M54 and Sgr are too faint (V$\sim$22) to be observed, thus 
the study of lower RGB stars represents the only possible route to infer 
$A(Li)_0$ in this galaxy.
Section~2 describes the spectroscopic observations, followed in Section~3 by 
the determination of the Li abundances and the constraints on $A(Li)_0$ for M54 stars, 
and is followed by a discussion of the results and conclusions.

\section{Observations}
\label{obs}

High-resolution spectra of lower RGB stars in M54 have been secured 
with the multi-object spectrograph FLAMES \citep{pasquini02} at the ESO Very Large Telescope, 
in the GIRAFFE/MEDUSA mode. The observations have been 
performed with the setups HR12 (to sample the Na D lines, with a resolution of 18700) 
and HR15N (to sample the Li doublet at 6707 \AA\ , with a resolution of 17000). 
The same target configuration has been used for both gratings and each target 
has been observed for a total time of 26 hr and 4 hr, for HR15N and HR12, 
respectively. 

\begin{figure}
\includegraphics[width=80mm]{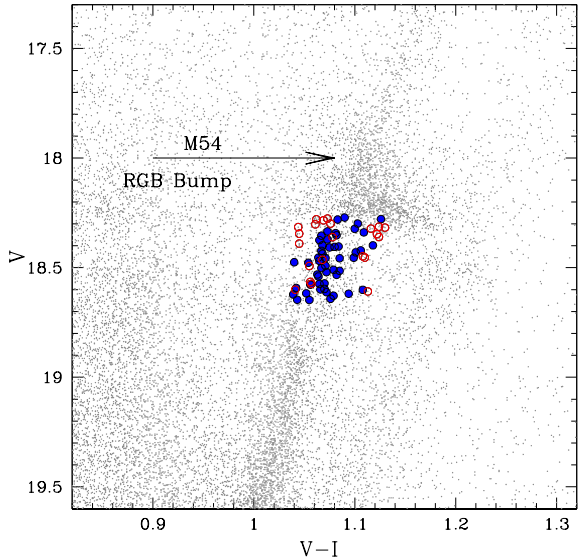}
\caption{
Colour-magnitude diagram of M54+Sgr that displays also 
the observed targets. Blue filled circles denote the 
member stars of M54, red empty circles the Sgr field stars.
}
\label{cmd}
\end{figure}

The targets have been selected from ACS@HST photometry \citep{siegel07} 
for the central region and from WFI@ESO photometry \citep{monaco02} for the 
outermost region. 
Eighty-five stars have been selected along the RGB of M54 in the magnitude range 
V=18.3-18.6, being its RGB bump at V$\sim$18, according to the RGB luminosity function. 
We excluded the 0.2 magnitudes below the RGB bump to minimise the contamination 
from the Sgr He-Clump stars. 
Figure~\ref{cmd} shows the colour-magnitude diagram of M54 with 
marked the observed targets (red and blue points).
The signal-to-noise (SNR) ratio per pixel around the Li doublet ranges 
from $\sim$30 to $\sim$50, with an average value of 42.

The spectra have been processed with the GIRAFFE data reduction 
pipeline, including bias-subtraction, flat-fielding, wavelength calibration, 
spectral extraction\footnote{https://www.eso.org/sci/software/pipelines/}.
Radial velocities have been measured with DAOSPEC \citep{stetson} 
by using $\sim$15 metallic lines. 11 targets have been discarded 
because they are clearly Galactic interlopers, with radial velocities 
between --105 and +60 km/s \citep[see Fig. 8 in][]{bellazzini08}.
Finally, our sample includes a total of 74 candidate member stars of M54 
(their main information is listed in Table 1).

\section{Chemical analysis}
\label{chem}

Values of $T_{eff}$ have been derived from the 
$(V-I)_0$ colour by means of the calibration by \citet{alonso99}, 
adopting the colour excess E(B-V)=~0.14 mag \citep{layden} 
and the extinction coefficients by \citet{mccall}. 
Surface gravities have been calculated from the Stefan-Boltzmann 
relation assuming the photometric $T_{eff}$, the bolometric 
corrections by \citet{alonso99} and the distance modulus $(m-M)_0$=~17.10 mag 
\citep{monaco04}. We assumed a mass of 0.8 $M_{\odot}$, according 
to a BaSTI isochrone \citep{pietr06} with 12 Gyr, Z=~0.0003
and $\alpha$-enhanced chemical mixture.
A microturbulent velocity $v_{turb}$=~1.5 km/s has been assumed 
for all the targets, taking the median value of $v_{turb}$ of the lower RGB stars 
analysed by MSB12.

\begin{figure}
\includegraphics[width=80mm]{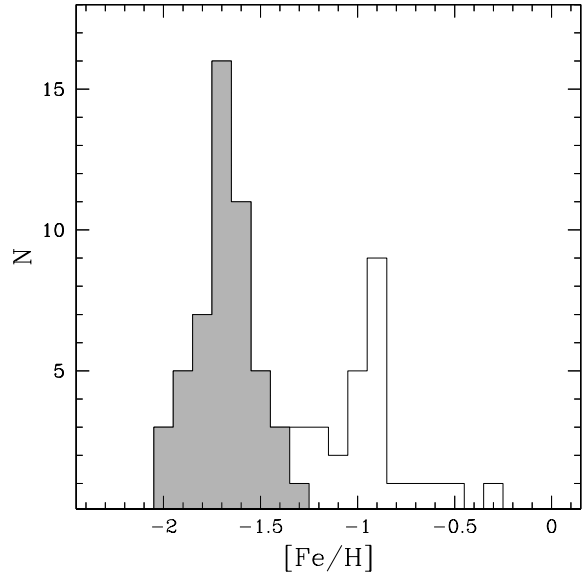}
\caption{[Fe/H] distribution for the RGB stars of M54. 
The grey shaded histogram includes the targets considered as members of M54, 
according to radial velocity and iron content.
}
\label{iron}
\end{figure}

Fe and Na abundances have been derived from the line equivalent widths (EWs)
by using the code GALA \citep{m13}, coupled with ATLAS9 model atmospheres.
Fe abundances have been obtained 
from the measure of $\sim$10-15 Fe~I lines, while Na abundances from the 
Na D lines at 5889-5895 \AA. 
EWs of Fe lines have been measured with DAOSPEC \citep{stetson}, 
while those of the Na lines by using IRAF assuming a Voigt profile. 
NLTE corrections for the Na abundances are from \citet{gratton99}.
The recent NLTE calculations by \citet{lind11} provide 
$[Na/Fe]_{NLTE}$ lower by about 0.2--0.3 dex; however, in the following we refer 
to the abundances obtained with the corrections by \citet{gratton99} 
to allow a direct comparison with \citet{carretta10} that measured Na abundances 
in 76 stars of M54.
Figure ~\ref{iron} shows the metallicity distribution of the 74 candidate M54 member stars, 
ranging from [Fe/H]=--2.0 dex up to --0.34 dex, with a main peak
at $\sim$--1.7 dex and a second peak at $\sim$--0.9 dex.

We consider as members of M54: {\sl (i)} stars with radial velocity between 100 and 170 km/s, 
\citep{bellazzini08}, and {\sl (ii)} stars with [Fe/H]$<$--1.3 dex, in order to exclude the stars 
of the second peak observed in the metallicity distribution, likely belonging to the Sgr field 
(note that the metallicity distributions of M54 by \citealt{bellazzini08} and 
\citealt{carretta10} are both broad but they do not show evidence of bimodality).
Finally, 51 targets are considered as {\sl bona fide} M54 member stars.
These stars are shown as blue 
circles in Fig.~\ref{cmd} and as the shaded histogram in Fig.~\ref{iron}. 
The mean iron content is [Fe/H]=~--1.67$\pm$0.02 dex ($\sigma$=~0.15 dex), 
compatible with those derived by \citet{bellazzini08} and \citet{carretta10}.
The M54 member stars show a wide range of [Na/Fe], between --0.56 and 
+0.77 dex, with an average value [Na/Fe]=+0.11$\pm$0.04 dex ($\sigma$=~0.31 dex), 
fully consistent with the results by \citet{carretta10}.

\begin{figure}
\includegraphics[width=80mm]{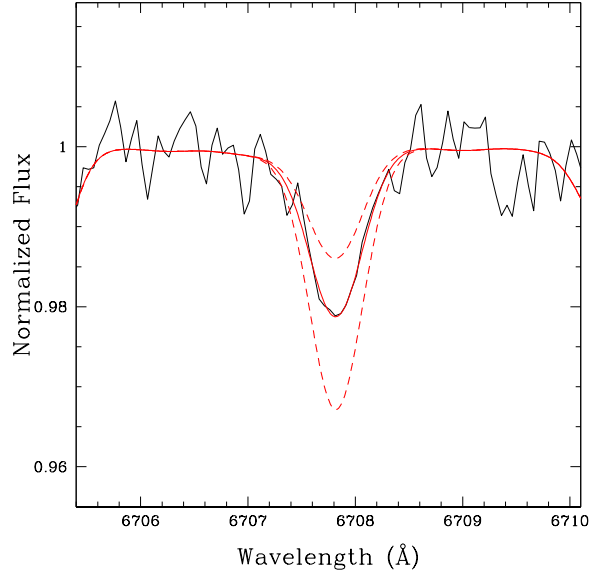}
\caption{The observed Li doublet of the average spectrum obtained by combining all 
51 targets that are members of M54. The red solid line is the best-fit synthetic spectrum, whilst 
the red dashed lines display the synthetic spectra calculated with $\pm$0.2 dex variations  
with respect to the best-fit abundance.}
\label{spec}
\end{figure}

The Li abundances have been derived from the Li resonance doublet at $\sim$6707 \AA, 
by comparing the observed spectra with a grid of synthetic spectra, calculated with 
the code SYNTHE \citep{sbordone04}. 
NLTE corrections are from \citet{lind08}.
The uncertainty in the fitting procedure has been estimated 
with MonteCarlo simulations performed by analysing synthetic 
spectra with the injection of Poissonian noise. Also, we included 
in the total error budget of the Li abundance the impact of the 
uncertainties in $T_{eff}$, 
the other parameters having a negligible impact on A(Li). 
Because of the weakness of the Li doublet (EW$\sim$13 m\AA\ ), 
at the SNR of our spectra it cannot be 
properly measured in each individual spectrum. Thus, we grouped together all 
the spectra of the stars considered as members of M54, 
obtaining an average spectrum with SNR$\sim$300 and 
assuming the average atmospheric parameters of the sample, 
namely $T_{eff}$=~4995 K and log~g=~2.46.
These stars are located in a narrow region of the colour-magnitude diagram, 
legitimating this procedure. In particular, $T_{eff}$ is the most critical 
parameter for the Li abundance estimate, whereas log~g and $v_{turb}$ have 
a negligible impact. The 51 cluster members cover  
a $T_{eff}$ range between 4873 K and 5090~K, with 
a mean equal to 4995 K ($\sigma$=~48~K), and a median value of 5005~K 
with an interquartile range of 51~K.
Figure~\ref{spec} shows the Li doublet observed in the average spectrum, with 
superimposed the best-fit synthetic spectrum (red solid line) and two synthetic spectra 
calculated with $\pm$0.2 dex with respect to the best-fit abundance (red dashed lines).

The final derived Li abundance is $A(Li)_{NLTE}$=~0.93$\pm$0.03$\pm$0.11~dex (where the 
first errorbar is the internal error as derived by the MonteCarlo simulations,  
and the second one is due to the $T_{eff}$ uncertainty). 
For consistency with MSB12 we checked also $A(Li)_{NLTE}$ obtained with 
the NLTE corrections by \citet{carlsson94}, that lead to an increase of 
the final abundance by 0.08~dex, thus providing $A(Li)_{NLTE}$=~1.01 dex. 
The choice of the NLTE corrections has obviously a small impact of the final A(Li) value 
and does not change drastically our conclusions.

\subsection{Checks about the average spectrum} 
\label{che}
 
To assess the stability of our results against the way we group the spectra, 
we have performed a number of sanity checks. 
In these tests we divided the cluster sample into two bins, according to: \\
{\sl (a)}~[Fe/H]; the two groups include stars 
with [Fe/H] lower and higher than the median value of [Fe/H]  
([Fe/H]=--1.67; see Fig.~\ref{iron}) respectively; \\
{\sl (b)}~$T_{eff}$; the boundary between the two groups is the median 
$T_{eff}$;\\
{\sl (c)}~magnitude; the two groups include stars 
fainter and brighter than the median V-band magnitude (V=~18.45) respectively;\\
{\sl (d)}~[Na/Fe]; the boundary between the two groups is  
the median value of the $[Na/Fe]_{NLTE}$ distribution ($[Na/Fe]_{NLTE}$=+0.16 dex).\\

\begin{figure}
\includegraphics[width=80mm]{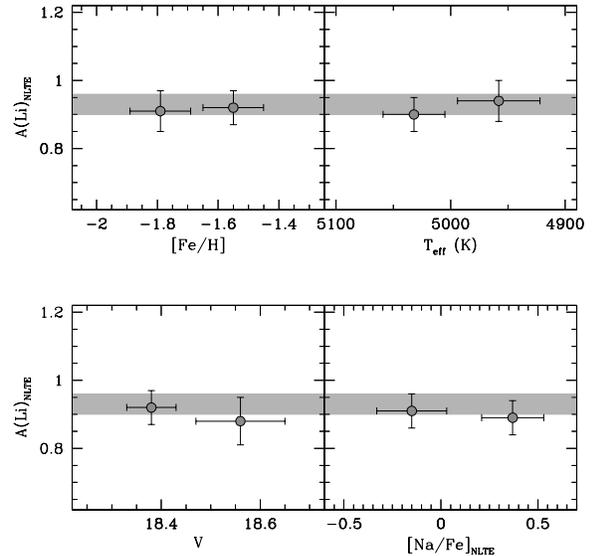}
\caption{
$A(Li)_{NLTE}$ values (dark grey circles) 
obtained by grouping the sample of M54 member stars 
into two average spectra according to the median value of [Fe/H] 
(left upper panel), $T_{eff}$ (right upper panel), 
V-band magnitude (left lower panel) and [Na/Fe] (right lower panel). 
Abundance errorbars include only the internal uncertainty from MonteCarlo 
simulations. Errorbars along the x-axis denote the 1$\sigma$ spread 
around the mean value of each quantity.
The shaded grey area in each panel denotes the $\pm1\sigma$ range with respect to the 
$A(Li)_{NLTE}$ obtained from the average spectrum of the whole M54 sample.
}
\label{gr}
\end{figure}

For all these cases, we found $A(Li)_{NLTE}$ 
compatible within the uncertainties with the value obtained with the average spectrum of the whole cluster targets, 
as shown by Fig.~\ref{gr}. 
The largest difference (0.08~dex, still compatible within 1$\sigma$ with the original 
value), is found when we group together spectra with V-band magnitude fainter than 
V=~18.45, because they have the lowest SNR.
In light of these results, 
we can conclude that no significant biases 
related to the grouping of the target spectra affect our Li abundance estimate.

Another point to discuss here concerns the use of a single value of the  NLTE correction  
computed for the average atmospheric parameters of the whole sample. 
To this purpose we notice that the variation of the NLTE corrections in the parameter space covered by 
our targets is small: in particular, at fixed $T_{eff}$/logg the corrections vary 
by $\sim$0.03--0.04 dex between the minimum and maximum [Fe/H] of the metallicity 
distribution, while at fixed metallicity, the corrections change by $\sim$0.03 dex 
between the minimum and maximum $T_{eff}$. 
To investigate more rigorously this effect, we simulated a spectrum with the following procedure:
(1)~for each individual member star a synthetic spectrum has been calculated 
with the appropriate atmospheric parameters and iron abundance, imposing a Li abundance 
$A(Li)_{NLTE}$=~0.93 dex (to take into account the proper NLTE correction of each star);
(2)~the spectra have been rescaled according to the relative differences 
in magnitude;
(3)~Poissonian noise has been injected in each synthetic spectrum to 
reproduce the measured SNR of the observed counterpart;
(4)~all these synthetic spectra have been co-added as done with the 
observed sample.

The entire procedure is repeated to obtain a sample of 1000 average 
spectra that has been analysed as done with the observed stars. 
The derived $A(Li)_{NLTE}$ distribution (assuming a single value of the NLTE correction) 
displays a mean value equal to 0.95~dex with a dispersion of 0.04~dex. This simulation 
confirms that star-to-star variations of the NLTE corrections are only a second order 
effect and do not affect substantially the abundance derived from the average spectrum.


\subsection{Lithium abundance and chemical anomalies in GCs}
\label{lian}

It is well established that individual GCs harbour 
sub-populations characterised by different 
abundances of light elements, like Na and O \citep[see e.g.][]{gratton12}. 
In principle, these so-called {\sl second-generation} stars, 
characterised by high values of [Na/Fe] and 
low values of [O/Fe], should 
display lower Li abundances, because they are predicted to be born 
from gas diluted with Li-poor material coming from asymptotic giant branch  or 
fast-rotating massive stars. 
Given that the thermonuclear reactions able to produce the observed chemical patterns  
occur at temperatures larger than $\sim10^7$K, while Li is destroyed at lower temperatures 
($\sim2.5\cdot10^6$K), second-generation stars should exhibit lower abundances of Li 
compared to first-generation stars. 
In particular, Li depletions, Li-O correlations and Li-Na anti-correlations are expected 
within individual clusters. 
Empirically, clear Li-O correlations have been detected in NGC6752 \citep{shen10} and 
47 Tuc \citep{dob}. 
Three Na-rich stars (thus belonging to the second cluster generation) with 
low Li abundance (A(Li)$<$2.0 dex) have been detected in NGC6397 \citep{lind09}, 
while most of the observed stars display a uniform Li (compatible with the 
{\sl Spite Plateau}) but a large range of Na, suggesting that 
Li depletion is negligible for the second generation stars of this cluster.
M4 displays a very small (if any) intrinsic Li dispersion, without correlation between 
O and Li abundance \citep{m11} and with a weak Li-Na anticorrelation \citep{monaco12}.
Lower RGB stars in M12 share all the same Li content, whilst there is a spread of Li in M5, but no  
statistically significant Li-O correlations and Li-Na anticorrelations \citep{dorazi14}.

We have checked whether potential 
systematic differences between A(Li) of first and second generation stars in M54 can affect our conclusions.
As discussed in Section~\ref{che} we divided the sample of M54 stars 
into two groups, according to their [Na/Fe] abundances, adopting as boundary the 
median value of the [Na/Fe] distribution (+0.16~dex). The derived average spectra 
show a very similar Li content, $A(Li)_{NLTE}$=~0.91$\pm$0.05 and 0.89$\pm$0.05 dex 
for the Na-poor and Na-rich groups, respectively, consistent with the value for the whole sample 
(see left bottom panel in Fig.~\ref{gr}).
Note that systematic differences in the Li content between the two samples smaller than $\sim$0.1 dex
(compatible, for instance, with those observed by \citealt{monaco12} in M4) cannot be ruled out.
However, such a small possible Li depletion in Na-rich stars of M54 does not change 
our conclusion about $A(Li)_0$ in this cluster.

\subsection{$A(Li)_0$ in M54}
To constrain the initial $A(Li)_0$ in M54, we adopted the same procedure 
discussed in MSB12, by using the amount of Li depletion due to the FDU
as predicted by stellar models (see their Table 2). 
For a metallicity [Fe/H]=--1.67 dex, the predicted value
is equal to 1.36~dex and 1.42~dex without and with atomic diffusion, respectively. 
As already discussed by MSB12, the amount of Li depletion 
along the RGB Plateau is marginally sensitive to the efficiency of the atomic diffusion 
that affects the dwarf stars much more strongly.
We recall that M54 has an intrinsic iron dispersion \citep{carretta10}; 
however, the predicted Li depletion changes by $\pm$0.02 dex with respect 
to the values quoted above if we consider the minimum and maximum value of the cluster 
metallicity distribution, namely [Fe/H]=--2.0 and --1.3 dex. 
We can thus neglect the effect of the cluster metallicity spread.

The derived $A(Li)_0$ in M54 is $A(Li)_0$=~2.29$\pm$0.11 dex (the error bar takes into account only the dominant effect 
of the uncertainty in $T_{eff}$)
without diffusion and 2.35$\pm$0.11 dex with fully efficient diffusion, 
When the NLTE corrections by \citet{carlsson94} are adopted, the range of $A(Li)_0$ 
values is 2.37--2.43 dex.

\section{Discussion} 

This is the first study of the primordial Li abundance 
in M54 and, hence, in the Sgr galaxy. Also, it is the most distant measurement of
$A(Li)$ in old, metal-poor stars obtained so far, given that 
Li abundance determinations in dwarf stars are restricted to distances within 
$\sim$8 kpc from the Sun \citep[see the case of M92,][]{boesgaard98,bonifacio02}. 
The use of lower RGB stars allows a giant leap in the 
study of $A(Li)_0$, pushing our investigation to $\sim$25 kpc from the Sun 
and enlarging our perspective of the Li problem. 
This work demonstrates the potential of lower RGB stars to investigate $A(Li)_0$ 
in stellar systems for which the observation of dwarf stars is precluded.

\begin{figure}
\includegraphics[width=80mm]{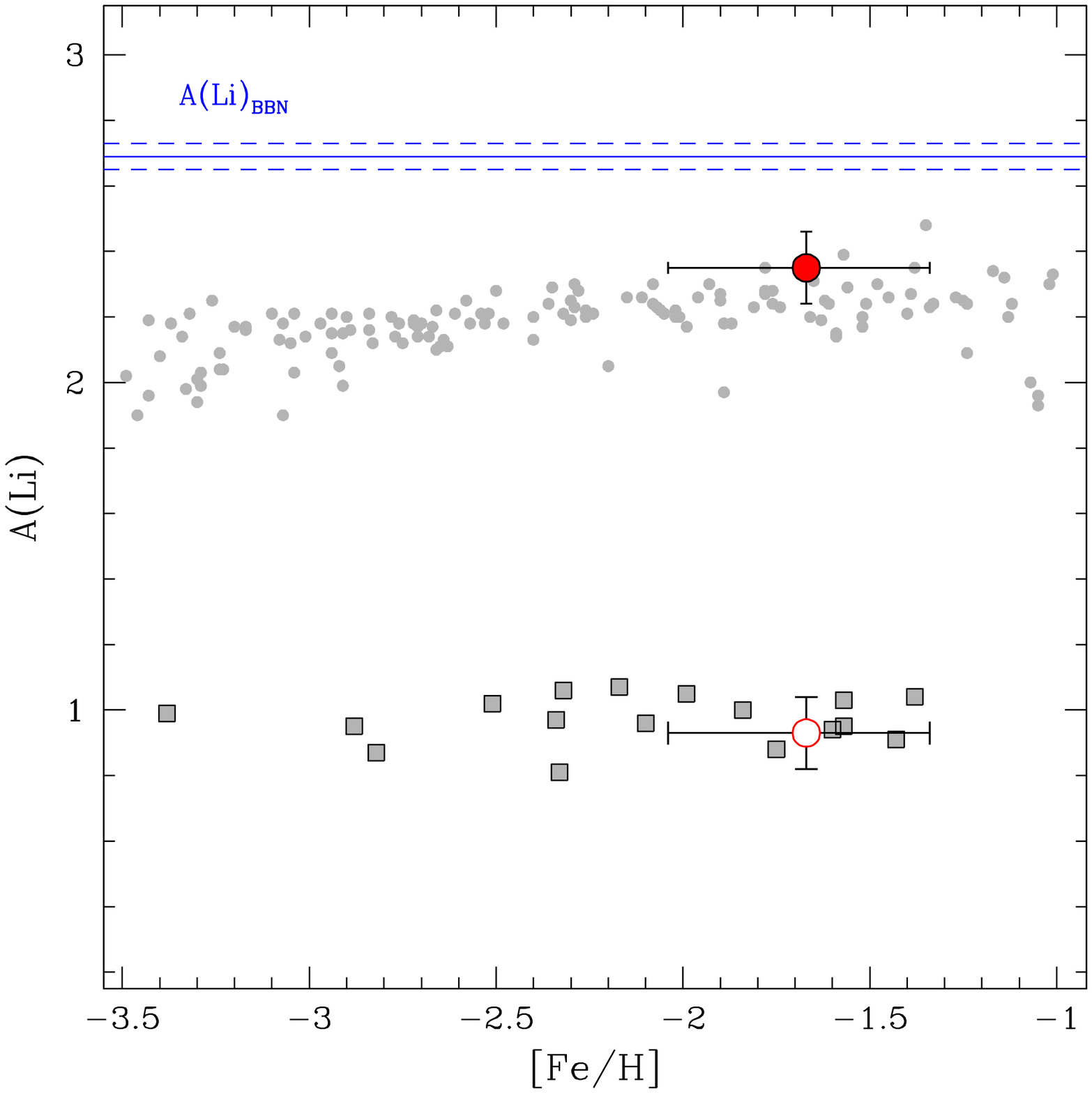}
\caption{Li abundance as a function of [Fe/H] for 
{\sl Spite Plateau} and lower RGB field halo stars. 
Grey circles denote the dwarf sample  
\citep{boni97,asplund06,aoki09,hosford09,melendez10}, and 
grey squares the lower RGB stars by MSB12. The empty red circle denotes  
the surface A(Li) in the lower RGB stars of M54, while the filled red 
circle displays the derived $A(Li)_0$ assuming fully efficient atomic diffusion 
(the horizontal errorbars associated to M54 data represent the 
range of [Fe/H] covered by the cluster). The blue solid line denotes $A(Li)_{BBN}$ \citep{coc13}, 
with the $\pm1\sigma$ uncertainty denoted by blue dashed lines.}
\label{lith}
\end{figure}

Fig.~\ref{lith} compares our $A(Li)$ and $A(Li)_0$ for M54 stars 
(red empty and filled circle, respectively)
to the results of Galactic field dwarf (grey circles) and lower RGB stars (grey squares).
The value of $A(Li)_{BBN}$ provided by \citet{coc13} is shown as reference.
First of all, $A(Li)$ measured in M54 red giants is in very good agreement with the results 
for the Galactic halo field 
(MSB12 found an average $A(Li)$=0.97 with the same $T_{eff}$ scale used for this study). 
Secondly, $A(Li)_0$ inferred from the lower RGB of M54 has, as already said, a very small 
dependence on whether atomic diffusion is fully efficient or inhibited, 
and results to be on average $\sim0.04-0.10$~dex higher than typical $A(Li)$ values 
measured in dwarf stars, that are equal on average to A(Li)$\sim$2.25 dex (see Fig.~\ref{lith}).
Assuming the initial Li in M54 and the Galactic halo was the same,  
if atomic diffusion is fully efficient in {\sl Spite Plateau} stars within the range of 
metallicities covered by M54 lower RGB stars, their surface Li abundances should be 0.4-0.7 dex
lower than $A(Li)_0$ \citep[see e.g. Fig.~3 in][]{m11}\footnote
{It is worth bearing in mind that a detailed comparison between $A(Li)_0$ derived from lower RGB stars 
and the {\sl Spite Plateau} depends also on the adopted $T_{eff}$ scales and NLTE corrections; here   
we simply take at face value the various estimates displayed in Fig.~\ref{lith}}. 

This means that either atomic diffusion is completely inhibited in halo field stars, 
and therefore the cosmological Li problem persists, or an additional element 
transport must be at work, burning during the main sequence more Li than predicted 
by models with diffusion only. This route has been investigated in order to 
interpret the surface Li abundances measured in dwarf stars of Galactic globular clusters.


To this purpose we first compare the results for M54 
with measurements of $A(Li)$ obtained for lower RGB stars in Galactic GCs that do not 
display a significant spread of Li. 
MSB12 determined $A(Li)$=1.00 and 
$A(Li)$=0.92 dex (both with $\sim$0.10~dex error bars) for  NGC6397 ([Fe/H]$\sim -2.1$ dex) and 
M4 ([Fe/H]$\sim -$1.1 dex) respectively, 
using the same $T_{eff}$ scale employed here. 
The same result has been found for lower RGB stars in M4 by \citet{villanova}.
These values are well consistent with M54 result. 
The recent study by \citet{dorazi14} found again a similar value, 
$A(Li)$=0.98 dex, with an error bar of $\sim$0.10~dex 
(using again the same $T_{eff}$ scale of this work) 
for lower RGB stars in M12, another cluster with essentially no Li spread 
amongst lower RGB objects, and [Fe/H] similar to M54. 

Measurements of $A(Li)$ in dwarfs stars have been performed in 
M92 \citep[][]{boesgaard98,bonifacio02}
NGC6397 \citep[][]{korn06,lind08,gh09,nordlander12}, 
NGC6752 \citep[][]{shen10,gruyters13,gruyters14}, 
M4 \citep[][]{m11,monaco12},
47~Tuc \citep[][]{dorazi,dob}. 
To these GCs, we add also Omega Centauri \citep[][]{monaco10}, 
a globular cluster-like stellar system characterized by a wide range of 
metallicities and probably ages, and usually thought as the stripped core of a 
dwarf galaxy. 
All these works found that dwarf GC stars display on average a Li content compatible 
with the {\sl Spite plateau}, confirming cosmological Li problem.
The works on NGC6397 and NGC6752 by \citet{gruyters13} and \citet{gruyters14} have however 
addressed this issue by considering as potential solution the combined effect of atomic diffusion and an 
hypothetical extra mixing process. In the following we will consider the recent analysis by \citet{gruyters14} of Li 
abundances in NGC6752, that has a [Fe/H] very close to the mean value of M54.  
These authors followed the same procedures applied to infer $A(Li)_0$ in 
NGC6397 \citep[see][for the latest work on this cluster]{nordlander12}. 
They measured the abundances of Li, and additional metals like 
Mg, Ca, Ti and Fe, in cluster stars from the main sequence turn off to the lower red giant branch, and 
compared the abundance trends along these evolutionary phases with results from stellar model calculations by 
\citet{richard02}. The observed trends could be matched only by models where the effect of diffusion was 
modulated by an additional mixing 
that in \citet{richard02} calculations is modeled as a diffusive process with diffusion coefficient $D_T$ 
chosen as 

\begin{equation}
D_T = 400 D_{He}(T_0) \left[\frac{\rho}{\rho(T_0)}\right]^{-3}
\label{eq1}
\end{equation}

where $D_{He}(T_0)$ is the atomic diffusion coefficient of He at a reference 
temperature $T_0$, and $\rho(T_0)$ is the density of the stellar model at the same temperature.
This is a somewhat ad-hoc prescription, with the proportionality constant $400 D_{He}(T_0)$, and the 
steep dependence on $\rho$ being essentially free parameters.
A justification for the choice of the steep dependence on $\rho$ stems from the need to restrict 
the efficiency of this mixing to a narrow region below the outer convection zone, 
as suggested by the solar beryllium abundance, believed to be essentially unaltered since the formation of the solar system.  
The temperature $T_0$ is also a free parameter, that determines the depth where this 
diffusive mixing is most effective.
It is important to remark that so far there has not been any attempt to test whether this mixing prescription can be associated to 
a well established physical process like, i.e., rotationally induced mixing. 
Assuming that the prescription in Eq.~\ref{eq1} is realistic, 
\citet{gruyters14} found that the free parameter $T_0$ has to be set to log($T_0$)=6.2 
to match the observed abundance trends for NGC6752, 
resulting in $A(Li)_0$=2.53$\pm$0.10, within less than 2$\sigma$ of the BBN predictions.

To our purposes it is relevant to notice that when log($T_0$)=6.2, the lower RGB abundances of \citet{richard05} models 
decrease by $\sim$0.1~dex compared to the case 
of pure diffusion, because during the main sequence additional Li is transported to the burning region by this extra mixing.
If the same process and the same efficiency estimated for NGC6752 are assumed also for M54, we need to add the same amount 
to $A(Li)_0$ determined including efficient diffusion, thus obtaining  
$A(Li)_0\sim$2.45$\pm$0.11~dex 
\citep[or $A(Li)_0\sim$2.53$\pm$0.11~dex when considering the NLTE corrections by ][]{carlsson94}. 

Given the current lack of identification of the proposed additional mixing with an established 
physical process, it is fair to say that   
we should be still cautious about this route to solve the cosmological Li problem, because 
simple parametric models have little predictive power. 
For example, to explain abundance trends in NGC6397, NGC6752 and M4 --and reconcile 
the measured $A(Li)$ with $A(Li)_{BBN}$-- one needs to employ a varying 
value of $T_0$, generally increasing with increasing [Fe/H]. Whether or not 
this trend of $T_0$ with [Fe/H] is a sign of the inadequacy of this parametrization of 
the additional mixing, requires a deeper understanding of its origin.

Observationally, \citet{gh09} found a trend of the surface A(Li) with $T_{eff}$ in NGC6397
that is not explainable with the additional mixing of Eq.~\ref{eq1}.
Also, as discussed by \citet{dob}, the constant Li abundance observed among the stars 
in Omega Centauri \citep{monaco10} spanning a wide range of ages and metallicities, and 
the Li distribution observed in 47~Tuc seem to require fine-tuned mechanisms that are at present difficult to explain 
with simple parametric diffusive mixing prescriptions.

\section{Conclusions} 

We measured the surface Li abundance in lower RGB stars harboured by M54, a GC belonging to the 
Sgr dwarf galaxy. 
We have obtained $A(Li)$=~0.93$\pm$0.11~dex, in agreement with 
measurements in Galactic halo stars. 
By considering the dilution due to the FDU, we have 
established an initial Li abundance of this stellar system ($A(Li)_0$=~2.29$\pm$0.11 
and 2.35$\pm$0.11 dex, without and with atomic diffusion, respectively) that is 
lower than the BBN value by $\sim$0.3~dex. 
The cluster $A(Li)_0$ can become compatible with $A(Li)_{BBN}$ within $\sim 2\sigma$ 
only assuming diffusion plus the additional mixing prescriptions by \citet{richard05} 
calibrated on the (same metallicity) Galactic GC NGC6752 \citep{gruyters14}.
Alternatively, inadequacies of the BBN model used to derive $A(Li)_{BBN}$ cannot be 
totally ruled out.

Also, an important question can be addressed by our study:
is the Li problem a {\sl local} problem, limited to our Galaxy, or is it independent 
of the environment? 
The analysis of the RGB stars in M54 confirms the findings in 
$\omega$ Centauri \citep{monaco10}, considered as the remnant of an accreted 
dwarf galaxy: {\sl the Li problem seems to be an universal problem, regardless 
of the parent galaxy}. The solution able to explain the discrepancy must work 
both in the Milky Way and other galaxies, with different origins and star formation histories. 
Thus, it seems unlikely that the scenario proposed by \citet{piau}, requiring that at least one third of 
the Galactic halo has been processed by Population III, massive stars, can work in the same way 
also in smaller systems like Sgr and $\omega$ Centauri \citep[see also][]{prantzos07}.
The universality of the {\sl Spite plateau} and the lower RGB abundances 
is a constraint that must be satisfied by any theory aimed at solving the 
cosmological Li problem.\\

We warmly thank the referee, Andreas Korn, for his detailed comments that have helped 
to improve the paper significantly.
S.V. gratefully acknowledges the support provided by Fondecyt reg. N. 1130721.

\begin{table*}
\begin{minipage}{125mm}
\caption{Identification numbers, coordinates, effective temperature, surface gravity, radial velocity, 
[Fe/H] and [Na/Fe] abundances. Final flag indicates the membership to M54 or to Sgr.} 
 \label{anymode}
\begin{tabular}{ccccccccc}
\hline
ID &  RA& Dec &   $T_{eff}$ & log~g & RV & [Fe/H]   & [Na/Fe] & flag \\
   & (J2000) & (J2000) & (K) &  & (km/s)  &  (dex)  & (dex) \\
\hline
\hline     
      6750      &   283.7948303     &  -30.4990501     &   5010      &  2.50	  &   130.5	 &     -1.75	  &   -0.56   &   M54	\\
      7590      &   283.7933655     &  -30.4935970     &   5010      &  2.51	  &   140.9	 &     -1.65	  &    0.38   &   M54	\\
     12291      &   283.7864380     &  -30.5019646     &   4987      &  2.47	  &   151.8	 &     -2.00	  &   -0.40   &   M54	\\
     21190      &   283.7769165     &  -30.5052319     &   4921      &  2.40	  &   135.7	 &     -1.74	  &    0.16   &   M54	\\
     51661      &   283.7530823     &  -30.5037270     &   4916      &  2.41	  &   148.5	 &     -0.34	  &   -0.48   &   Sgr	\\
     53985      &   283.7514343     &  -30.4937611     &   4936      &  2.37	  &   149.1	 &     -1.74	  &    0.09   &   M54	\\
     56686      &   283.7486572     &  -30.5045719     &   5018      &  2.45	  &   141.6	 &     -1.42	  &    0.25   &   M54	\\
     65022      &   283.7391968     &  -30.5047073     &   5046      &  2.52	  &   149.3	 &     -1.24	  &    0.46   &   Sgr	\\
     69373      &   283.7333679     &  -30.4932117     &   4977      &  2.37	  &   144.5	 &     -1.56	  &   -0.21   &   M54	\\
     75429      &   283.7950134     &  -30.4848213     &   5028      &  2.50	  &   149.3	 &     -1.58	  &   -0.19   &   M54	\\
     86412      &   283.7847900     &  -30.4922523     &   5079      &  2.56	  &   146.1	 &     -1.95	  &    0.58   &   M54	\\
     91967      &   283.7814636     &  -30.4800529     &   4975      &  2.42	  &   142.1	 &     -1.86	  &    0.47   &   M54	\\
    121249      &   283.7685242     &  -30.4917202     &   4873      &  2.32	  &   153.5	 &     -1.86	  &   -0.07   &   M54	\\
    141357      &   283.7619019     &  -30.4915009     &   5023      &  2.51	  &   146.1	 &     -1.87	  &    0.25   &   M54	\\
    155785      &   283.7575684     &  -30.4852924     &   5023      &  2.43	  &   141.1	 &     -1.66	  &    0.26   &   M54	\\
    201571      &   283.7276917     &  -30.4915905     &   5015      &  2.46	  &   144.4	 &     -0.96	  &    0.04   &   Sgr	\\
    208256      &   283.7915344     &  -30.4739513     &   5082      &  2.54	  &   143.7	 &     -1.93	  &    0.16   &   M54	\\
    216867      &   283.7841797     &  -30.4684467     &   5007      &  2.47	  &   142.7	 &     -1.79	  &    0.23   &   M54	\\
    231677      &   283.7753906     &  -30.4778271     &   5048      &  2.55	  &   136.0	 &     -1.70	  &   -0.22   &   M54	\\
    235280      &   283.7738342     &  -30.4735279     &   5056      &  2.54	  &   147.3	 &     -1.66	  &    0.53   &   M54	\\
    279832      &   283.7575073     &  -30.4666042     &   5090      &  2.56	  &   148.6	 &     -1.79	  &    0.58   &   M54	\\
    299467      &   283.7481689     &  -30.4728985     &   4960      &  2.36	  &   149.6	 &     -1.72	  &    0.27   &   M54	\\
    304691      &   283.7450256     &  -30.4665394     &   5005      &  2.48	  &   139.0	 &     -1.31	  &    0.18   &   M54	\\
    315861      &   283.7359009     &  -30.4745407     &   5025      &  2.50	  &   140.6	 &     -1.60	  &   -0.07   &   M54	\\
    335718      &   283.7800903     &  -30.4574833     &   5005      &  2.52	  &   150.1	 &     -1.67	  &   -0.28   &   M54	\\
    340297      &   283.7754211     &  -30.4570541     &   4980      &  2.47	  &   136.6	 &     -1.63	  &   -0.14   &   M54	\\
    342644      &   283.7732849     &  -30.4543114     &   5018      &  2.42	  &   142.6	 &     -1.62	  &    0.10   &   M54	\\
    348795      &   283.7681274     &  -30.4567890     &   5002      &  2.41	  &   149.6	 &     -1.73	  &    0.30   &   M54	\\
    356601      &   283.7614441     &  -30.4637871     &   5025      &  2.46	  &   145.1	 &     -1.59	  &    0.53   &   M54	\\
    358028      &   283.7607117     &  -30.4514027     &   4928      &  2.36	  &   144.2	 &     -1.66	  &    0.01   &   M54	\\
    359389      &   283.7593689     &  -30.4573898     &   5051      &  2.48	  &   147.2	 &     -1.37	  &   -0.14   &   M54	\\
    379953      &   283.7375183     &  -30.4624958     &   5048      &  2.49	  &   137.9	 &     -0.90	  &   -0.14   &   Sgr	\\
   1031659      &   283.7920227     &  -30.4277306     &   5002      &  2.38	  &   161.6	 &     -1.15	  &   -0.07   &   Sgr	\\
   1031785      &   283.7452393     &  -30.5135555     &   5030      &  2.40	  &   136.6	 &     -1.12	  &    0.03   &   Sgr	\\
   1032003      &   283.8471985     &  -30.3341923     &   5012      &  2.39	  &   140.1	 &     -0.71	  &    0.21   &   Sgr	\\
   1032576      &   283.7683716     &  -30.4011650     &   4995      &  2.39	  &   177.8	 &     -1.25	  &   -0.29   &   Sgr	\\
   1032677      &   283.6716309     &  -30.3297195     &   5033      &  2.41	  &   151.0	 &     -0.86	  &   -0.15   &   Sgr	\\
   1033129      &   283.5857239     &  -30.4534187     &   4878      &  2.34	  &   150.6	 &     -0.79	  &     ---   &   Sgr	\\
   1033207      &   283.7309570     &  -30.5093040     &   5077      &  2.43	  &   137.6	 &     -0.90	  &   -0.50   &   Sgr	\\
   1033253      &   283.7126770     &  -30.3716583     &   4864      &  2.33	  &   145.8	 &     -0.88	  &   -0.71   &   Sgr	\\
   1033431      &   283.7608337     &  -30.6025276     &   4897      &  2.35	  &   165.9	 &     -1.14	  &   -0.05   &   Sgr	\\
   1033794      &   283.8335571     &  -30.6086063     &   4953      &  2.38	  &   142.7	 &       ---	  &     ---   &   Sgr	\\
   1033808      &   283.6697998     &  -30.5691261     &   4975      &  2.39	  &   144.7	 &     -0.91	  &    0.03   &   Sgr	\\
   1034001      &   283.7369385     &  -30.4347324     &   4914      &  2.37	  &   146.6	 &     -1.68	  &   -0.48   &   M54	\\
   1034068      &   283.7054138     &  -30.4942036     &   4982      &  2.40	  &   141.2	 &     -1.67	  &    0.54   &   M54	\\
   1034166      &   283.6147766     &  -30.5109158     &   5074      &  2.44	  &   102.8	 &     -0.95	  &    0.44   &   Sgr	\\ 
   1034215      &   283.6256104     &  -30.4640865     &   4883      &  2.35	  &   162.6	 &     -0.56	  &   -0.49   &   Sgr	\\
   1034363      &   283.7250061     &  -30.4443989     &   4980      &  2.40	  &   146.5	 &     -1.48	  &    0.24   &   M54	\\
   1034592      &   283.8386841     &  -30.4815445     &   4878      &  2.36	  &   159.7	 &     -0.47	  &   -0.68   &   Sgr	\\
   1034628      &   283.8795471     &  -30.3539162     &   4990      &  2.41	  &   144.3	 &     -1.00	  &   -0.62   &   Sgr	\\
   1034807      &   283.7220154     &  -30.3370037     &   4627      &  2.32	  &   147.1	 &     -0.46	  &   -0.49   &   Sgr	\\ 
   1034871      &   283.6983032     &  -30.4932556     &   5002      &  2.42	  &   148.2	 &     -1.74	  &    0.77   &   M54	\\
   1035051      &   283.5896912     &  -30.4579124     &   4975      &  2.41	  &   149.7	 &     -0.94	  &   -0.98   &   Sgr	\\
   1035061      &   283.8948059     &  -30.4827633     &   4678      &  2.36	  &   155.8	 &     -0.83	  &   -0.55   &   Sgr	\\
   1035450      &   283.7792969     &  -30.5218792     &   5074      &  2.46	  &   142.8	 &     -0.94	  &    0.33   &   Sgr	\\
   1035614      &   283.6965637     &  -30.4720631     &   4706      &  2.38	  &   141.5	 &     -0.55	  &   -0.71   &   Sgr	\\
   1035639      &   283.9363708     &  -30.3593540     &   4777      &  2.42	  &   138.1	 &     -0.55	  &     ---   &   Sgr	\\ 
   1035659      &   283.8937683     &  -30.5231647     &   5015      &  2.44	  &   143.6	 &     -1.77	  &   -0.15   &   M54	\\
   1035689      &   283.6646118     &  -30.4083195     &   4892      &  2.38	  &   125.6	 &     -1.75	  &    0.24   &   M54	\\
 \hline
\end{tabular}
\end{minipage}
\end{table*}

\begin{table*}
\begin{minipage}{125mm}
\contcaption{Identification numbers, coordinates, effective temperature, surface gravity, radial velocity, 
[Fe/H] and [Na/Fe] abundances. Final flag indicates the membership to M54 or to Sgr.} 
 \label{anymode}
\begin{tabular}{ccccccccc}
\hline
ID &  RA& Dec &   $T_{eff}$ & log~g & RV & [Fe/H]   & [Na/Fe] & flag \\
   & (J2000) & (J2000) & (K) &  & (km/s)  &  (dex)  & (dex) \\
\hline
\hline     
   1035733      &   283.6871338     &  -30.5677834     &   5074      &  2.46	  &   140.8	 &     -0.90	  &    0.23   &   Sgr	\\ 
   1035834      &   283.7239990     &  -30.4563236     &   4985      &  2.43	  &   141.3	 &     -1.55	  &   -0.27   &   M54	\\
   1035938      &   283.7536316     &  -30.4476051     &   4997      &  2.43	  &   123.1	 &     -1.74	  &   -0.32   &   M54	\\
   1035965      &   283.6389771     &  -30.5077534     &   4670      &  2.36	  &   147.7	 &     -0.39	  &     ---   &   Sgr	\\ 
   1036018      &   283.7001953     &  -30.5760975     &   4738      &  2.40	  &   140.0	 &	0.07	  &   -0.40   &   Sgr	\\ 
   1036558      &   283.7184753     &  -30.4782505     &   4933      &  2.41	  &   138.1	 &     -1.66	  &   -0.03   &   M54	\\
   1036741      &   283.7993469     &  -30.4916420     &   5015      &  2.45	  &   137.5	 &     -1.66	  &    0.06   &   M54	\\
   1036890      &   283.8543091     &  -30.5144444     &   4716      &  2.40	  &   116.0	 &     -1.04	  &   -0.24   &   Sgr	\\ 
   1037256      &   283.9817505     &  -30.4822559     &   4660      &  2.38	  &   146.0	 &     -0.56	  &   -0.43   &   Sgr	\\ 
   1037298      &   283.7445984     &  -30.5185242     &   4911      &  2.41	  &   158.1	 &     -0.96	  &    0.15   &   Sgr	\\
   1037347      &   283.5868835     &  -30.5343914     &   4747      &  2.43	  &   151.9	 &	0.24	  &    0.15   &   Sgr	\\ 
   1037357      &   283.7287598     &  -30.3941364     &   4938      &  2.42	  &   144.0	 &     -1.55	  &    0.27   &   M54	\\
   1037383      &   283.7582397     &  -30.4474907     &   5007      &  2.46	  &   151.0	 &     -1.58	  &    0.38   &   M54	\\
   1037405      &   283.8029785     &  -30.6097832     &   4972      &  2.44	  &   145.7	 &     -1.40	  &     ---   &   M54	\\ 
   1037499      &   283.9023132     &  -30.5804310     &   5082      &  2.49	  &   132.9	 &     -1.15	  &     ---   &   Sgr	\\
   1037755      &   283.8068848     &  -30.4766140     &   5023      &  2.47	  &   143.1	 &     -1.50	  &    0.27   &   M54	\\
   1037842      &   283.6522827     &  -30.4114075     &   4773      &  2.45	  &   149.4	 &     -0.84	  &   -0.16   &   Sgr	\\ 
   1037956      &   283.7883301     &  -30.5176754     &   5087      &  2.50	  &   129.4	 &     -1.62	  &    0.51   &   M54	\\
   1038371      &   283.7473450     &  -30.4219704     &   4687      &  2.41	  &   130.4	 &     -0.82	  &   -0.78   &   Sgr	\\ 
   1038827      &   283.7271729     &  -30.4614105     &   5018      &  2.48	  &   145.7	 &     -1.64	  &    0.07   &   M54	\\
   1038900      &   283.9519348     &  -30.5770645     &   4682      &  2.41	  &   143.3	 &     -0.30	  &   -0.47   &   Sgr	\\ 
   1039247      &   283.7293701     &  -30.5351334     &   4972      &  2.46	  &   146.1	 &     -1.52	  &    0.34   &   M54	\\
   1039380      &   283.6416626     &  -30.4188614     &   4764      &  2.46	  &   123.5	 &     -0.65	  &   -0.26   &   Sgr	\\ 
   1039482      &   283.9963989     &  -30.4818535     &   4782      &  2.47	  &   138.4	 &     -0.64	  &     ---   &   Sgr	\\ 
   1039645      &   283.7586670     &  -30.5772209     &   4800      &  2.48	  &   133.7	 &     -0.27	  &   -0.47   &   Sgr	\\
   1040277      &   283.8876648     &  -30.4286728     &   4807      &  2.49	  &   142.9	 &     -0.61	  &   -0.49   &   Sgr	\\
   1040695      &   283.8935852     &  -30.6129131     &   4775      &  2.48	  &   159.6	 &     -0.10	  &     ---   &   Sgr	\\ 
   1040996      &   283.7201233     &  -30.5874443     &   4716      &  2.45	  &   143.2	 &     -0.47	  &   -0.87   &   Sgr	\\ 
   1041212      &   283.7241211     &  -30.5361309     &   5043      &  2.52	  &   144.3	 &     -1.03	  &   -0.09   &   Sgr	\\
   1041214      &   284.0014343     &  -30.5138874     &   4816      &  2.51	  &   142.7	 &     -0.06	  &     ---   &   Sgr	\\ 
   1041231      &   283.7814636     &  -30.6557369     &   4682      &  2.44	  &   140.2	 &       ---	  &     ---   &   Sgr	\\ 
   1041308      &   283.8000793     &  -30.4399662     &   5046      &  2.52	  &   149.8	 &     -1.75	  &    0.06   &   M54	\\
   1041392      &   283.7596741     &  -30.6370296     &   4798      &  2.50	  &   160.3	 &	0.52	  &   -0.78   &   Sgr	\\ 
   1041896      &   283.8808899     &  -30.4472752     &   4718      &  2.47	  &   133.9	 &     -0.66	  &   -0.52   &   Sgr	\\
   1041905      &   283.8525391     &  -30.4917564     &   4784      &  2.50	  &   146.9	 &     -0.61	  &   -0.56   &   Sgr	\\ 
   1042086      &   283.8047180     &  -30.4664555     &   5020      &  2.52	  &   142.2	 &     -1.99	  &    0.59   &   M54	\\
   1042102      &   283.7253723     &  -30.4827061     &   5085      &  2.55	  &   143.9	 &     -0.92	  &     ---   &   Sgr	\\
   1042123      &   283.7680359     &  -30.4397469     &   4916      &  2.47	  &   141.1	 &     -1.77	  &   -0.11   &   M54	\\
   1042352      &   283.6452942     &  -30.4793129     &   4904      &  2.47	  &   165.3	 &     -1.25	  &   -0.03   &   Sgr	\\
   1042739      &   283.7419739     &  -30.4234066     &   4950      &  2.49	  &   146.3	 &     -1.47	  &   -0.22   &   M54	\\
   1043020      &   283.7697144     &  -30.5278034     &   4987      &  2.52	  &   141.2	 &     -1.87	  &     ---   &   M54	\\ 
   1043447      &   283.7137451     &  -30.3280296     &   4995      &  2.52	  &   154.8	 &     -1.51	  &   -0.21   &   M54	\\
\hline
\end{tabular}
\end{minipage}
\end{table*}

\end{document}